\documentclass[a4paper,aps,twocolumn]{revtex4}

\begin{document}

\title{On the Geometry and Entropy of Non-Hamiltonian Phase Space}

\author{Alessandro Sergi\footnote{E-mail: asergi@unime.it}
and Paolo V. Giaquinta
} \affiliation{ Dipartimento di Fisica,
Universit\'a degli Studi di Messina, Contrada Papardo
98166 Messina, Italy }

\begin{abstract}
We analyze the equilibrium statistical mechanics of canonical,
non-canonical and non-Hamiltonian equations of motion 
by throwing light into the peculiar geometric structure of phase
space. Some fundamental issues regarding time translation and phase space measure
are clarified.
In particular, we emphasize that a phase space measure should
be defined by means of the Jacobian
of the transformation between different types of coordinates 
since such a determinant is different from zero in the non-canonical case
even if the phase space compressibility is null.
Instead, the Jacobian determinant associated with phase space flows is unity
whenever non-canonical coordinates lead to a vanishing compressibility,
so that its use in order to define a measure may not be always correct.
To better illustrate this point, we derive a mathematical condition for defining
non-Hamiltonian phase space flows with zero compressibility.
The Jacobian determinant
associated with time evolution in phase space is altogether useful for analyzing
time translation invariance.
The proper definition of a phase space measure is particularly important
when defining  the entropy functional in the canonical, non-canonical,
and non-Hamiltonian cases.
We show how the use of relative entropies can circumvent
some subtle problems that are encountered when dealing with
continuous probability distributions and phase space measures.
Finally, a maximum (relative) entropy principle is formulated
for non-canonical and non-Hamiltonian phase space flows.
\end{abstract}

\maketitle


\today

\section{Introduction}

In this paper we address some general issues concerning
the geometry of non-Hamiltonian phase space under the condition of
thermodynamical equilibrium. 
In the field of classical molecular dynamics simulations,
non-Hamiltonian formalisms have been developed
in order to simulate numerically the effect 
of thermal and pressure baths
on relevant subsystems 
by means of a finite number of degrees
of freedom~\cite{andersen,nose,hoover,ferrario}.
It is also worth mentioning that the non-Hamiltonian approach is the method of choice
for path integral~\cite{pimd} and 
ab-initio path integral molecular dynamics calculations~\cite{aipimd}).
More recently, it has been shown that the theories~\cite{kakkacicco,bsiluranti}
needed to describe in a  consistent way the coupling between quantum 
and classical degrees of freedom
are non-Hamiltonian in their very essence~\cite{nielsen,ale}.
Despite the ever growing number of successful applications
of non-Hamiltonian theories to molecular dynamics simulations,
a clarification is desirable
since classical non-Hamiltonian theories in phase space have been typically
formulated by means of two distinct approaches.
In one case, phase space is considered as
a Riemann manifold, endowed with a metric tensor whose
determinant is used to define the measure of the volume
element~\cite{tuck,tarasov}.
Within this type of approach it is not clear how to derive
non-Hamiltonian equations of motion but, once these are given,
their statistical mechanics is defined by introducing an
invariant measure of phase space.
As a matter of fact, 
the \emph{metric} of phase space has been exclusively linked in Refs.~\cite{tuck,tarasov}
to the phase space compressibility. However,
there are systems 
that, while being non-canonical, have a non-zero Jacobian
and a vanishing compressibility.
To show this, we derive a condition
in order to define non-canonical and non-Hamiltonian systems
with zero compressibility.
Then, we argue that the phase space measure should be defined in terms of
the Jacobian pertaining to the transformation
between different types of coordinates (for example, canonical
\emph{vs.} non-canonical coordinates). 
This Jacobian should not be confused with another Jacobian determinant,
that is associated with the transformation realized 
by the law of motion, 
and which arises naturally when addressing the time-translation property
of statistical mechanics in phase space.

Another approach to non-Hamiltonian statistical mechanics
has been proposed in Refs.~\cite{b1-2,bdispettose}. It uses generalized
antisymmetric brackets to define non-Hamiltonian equations of motion.
A generalized Liouville equation for distribution
functions in phase space can be written in such a way that
its solutions naturally provide the statistical weight of phase space.
In addition, linear response theory~\cite{b1-2,bdispettose}
as well as extensions of the theory to quantum and
quantum-classical mechanics~\cite{bsiluranti} can be easily formulated.
Such an approach has an algebraic structure
whose distinctive feature is the violation of the 
Jacobi relation~\cite{goldstein,mccauley,arnold,morrison}.
It is important to note that the failure of the Jacobi relation
implies the lack of time-translation invariance of the algebra~\cite{nielsen}.
This means that if two arbitrary phase space observables 
obey a relation expressed by means of a generalized bracket at time $t_0$,
then, in general, such two observables 
will \underline{not} satisfy the same relation at time $t\neq t_0$.
This leads to consequences in the proper definition of the constants of motion
(for example, the generalized bracket of two constants of motion is not
necessarily a constant of motion) and in the definition of correlation functions.
An example of the application of a similar philosophy
to address non-Hamiltonian phase space can be found
in Ezra's work~\cite{ezra,ezra2}, where the elegant formalism
of differential forms and exterior derivatives has been used.
More recently, such an approach has been re-expressed
using the antisymmetric matrix structure introduced
in Refs.~\cite{b1-2,bdispettose}
in order to devise measure-preserving algorithms
for the numerical integration of non-Hamiltonian equations of motion~\cite{ezra3}.

Another issue that is particularly relevant for non-Hamiltonian statistical 
mechanics is the covariant definition of the entropy functional.
When dealing with non-canonical and non-Hamiltonian systems 
with zero compressibility, care must be payed
because, as discussed above, the conclusion that the measure 
is trivial may be erroneous~\cite{tuck,tarasov}.
Instead, the phase space Jacobian, which is different from unity
whenever non-canonical coordinates are adopted,
should be used to define the correct measure
of phase space and the covariant entropy functional.
However, such a functional,
as already discussed in Refs.~\cite{stratonovich,ramshaw,rau}, can be defined
without any \emph{a priori} knowledge of the \emph{metric} factor.
The key point is that relative entropies are called for when dealing
with continuous probability distributions.
In fact, a relative entropy
functional naturally accounts for unknown \emph{metric} factors.
Therefore, a maximum entropy principle can be introduced 
for non-canonical and non-Hamiltonian dynamics. 

Since the (relative) entropy production is trivially null
under the condition of thermodynamical equilibrium, a
comment is called for.
Indeed, it is worth underlining that, while presenting results
that aim at clarifying the nature of non-Hamiltonian equilibrium
geometry, we also intend to set the pre-conditions to cope
with the non-equilibrium case within an information theoretical perspective
(for a general introduction see Ref.~\cite{jaynes-prob} and references contained therein).
Within such a framework, it is not really important to describe
the fine-grained complexity of phase space under non-equilibrium
conditions: what matters is the ability to identify the statistical \emph{constraints}
which allow one to perform calculations with a predictive value.

This approach is different from another approach
to non-equilibrium statistical mechanics~\cite{neqham}
which, instead, aims at describing in a complete way
the fine-grained complexity of phase space
out of equilibrium. This latter approach has led, in recent years,
to some important achievements such as
the discovery of the fractal nature of phase space 
in non-equilibrium steady states~\cite{fractal} and an analysis of
some interesting statistical models such as Anasov's systems~\cite{ruelle}.
However, these systems are different from those investigated by 
chemical physics (see \cite{chemphys} for some recent examples).
However, we believe that looking at the fine-grained nature of phase space or
invoking information theoretic techniques
should not be considered as mutually exclusive paths of investigation.

Clarifying the above issues is not the only goal of this paper.
For an easier reading of the paper, we anticipate the results
we shall illustrate in the following:
\begin{itemize}
\item[(i)] We derive the conditions for obtaining 
a class of non-Hamiltonian phase-space flows with zero compressibility.

\item[(ii)] Given that the compressibility is not a critical feature of non-Hamiltonian phase-space flows, we introduce a more proper definition of the phase-space measure by means of the Jacobian of the coordinate
transformation.

\item[(iii)] We show that using the Jacobian determinant ${\cal J}(t,t_0)$ 
that is associated with the equations of motion 
in order to define the phase-space measure can be misleading when the 
compressibility vanishes. We also show that this determinant is relevant 
for the invariance properties of the statistical-mechanical averages with 
respect to time translation.

\item[(iv)] We introduce a maximum-entropy approach for non-Hamiltonian systems.

\item[(v)] We adopt the concept of relative entropy in order 
to ensure covariance in phase space. As a by-product of this analysis, we prove that 
Ramshaw's covariant entropy~\cite{ramshaw} is actually a relative entropy. 
\end{itemize}

The paper is divided into two main sections. 

Section~\ref{sec:geometry} treats the peculiar geometry
of phase space and its influence on dynamics 
and  statistical mechanics.
In Sec.~\ref{sec:canham},
the canonical Hamiltonian phase space is briefly reviewed.
The geometry of canonical phase space and
its statistical mechanics are shortly discussed
by means of Poisson brackets.
The invariance of equilibrium statistical mechanics
under time translation is discussed.
Section~\ref{sec:noncanham} generalizes the result of Sec.~\ref{sec:canham}
to non-canonical Hamiltonian phase space flows
by employing non-canonical brackets.
Section~\ref{sec:nonham} shows that non-Hamiltonian 
statistical mechanics is not time-translation invariant.
Conditions in order to obtain non-Hamiltonian phase space flows
with zero compressibility are also derived.
Appendices~\ref{app:1} and~\ref{app:2} provide examples
of simple non-canonical and non-Hamiltonian dynamics
with zero compressibility.

Section~\ref{sec:relativeentropy} is devoted to a discussion of the
relative entropy functional, of its covariance, and of a formula
for its production rate.
In Sec.~\ref{sec:hamentropy}
the covariant relative entropy is illustrated and is used in order
to derive a ``maximum relative entropy'' principle.
Such results are extended to the non-canonical and non-Hamiltonian cases
in Sec.~\ref{sec:ncrelentropy}.
A formula for the rate of production of the relative entropy
is derived in Sec.~\ref{sec:enprod}.
Concluding remarks are given in Sec.~\ref{sec:conclusions}.

\section{Geometry of phase space}\label{sec:geometry}

\subsection{Canonical Hamiltonian Phase Space}
\label{sec:canham}

Let $x=(q,p)$ denote a point in phase space, where $q$ and
$p$ are generalized coordinates and momenta, respectively.
Let ${\cal H}(x)$ be the (Hamiltonian) generalized energy function
of the system. We consider, here and in the following
section, only the case in which ${\cal H}(x)$ is time-independent.
Let $2n$ be the dimension of phase space. Then, the
$2n\times 2n$ antisymmetric matrix $\mbox{\boldmath$\cal B$}^s$,
or cosymplectic form~\cite{goldstein,mccauley,arnold,morrison},
reads as
\begin{equation}
\mbox{\boldmath$\cal B$}^s=
\left[\begin{array}{cc}{\bf 0} &{\bf 1}\\-{\bf 1} & {\bf 0}\end{array}\right]
\;.\label{b^s}
\end{equation}
Correspondingly, the canonical Hamiltonian equations of motion are given by
\begin{equation}
\dot{x}_i=\sum_{j=1}^{2n}{\cal B}_{ij}^s\frac{\partial{\cal H}}{\partial x_j}
\;,\label{canhameq}
\end{equation}
for $i=1,\ldots,2n$.
Equations~(\ref{canhameq}) can be derived from a variational principle in phase
space~\cite{pericolanti} applied to the action
written in symplectic form
\begin{equation}
{\cal A}=\int dt\left[\sum_{i,j=1}^{2n}\frac{1}{2}
\dot{x}_i{\cal B}_{ij}^sx_j-{\cal H}\right]\;.
\label{symplaction}
\end{equation}
The compressibility of phase space 
\begin{eqnarray}
\kappa&=&\sum_{i=1}^{2n}\frac{\partial\dot{x}_i}{\partial x_i}
=\sum_{i,j=1}^{2n}
\frac{\partial{\cal B}_{ij}^s}{\partial x_i}
\frac{\partial{\cal H}}{\partial x_j}
\end{eqnarray}
is zero because
the  matrix $\mbox{\boldmath$\cal B$}^s$ is constant.
Poisson brackets can be defined as
\begin{equation}
\{a(x),b(x)\}_{\mbox{\tiny\boldmath$\cal B$}^s}=\sum_{i,j=1}^{2n}
\frac{\partial a}{\partial x_i}{\cal B}_{ij}^s
\frac{\partial b}{\partial x_j}\;,
\end{equation}
where $a(x)$ and $b(x)$ are arbitrary phase space functions,
so that the equations of motion follow in the form
\begin{equation}
\dot{x}_i=\{x_i,{\cal H}\}_{\mbox{\tiny\boldmath$\cal B$}^s}\;.
\end{equation}
The Jacobi relation
\begin{equation}
\{a,\{b,c\}_{\mbox{\tiny\boldmath$\cal B$}^s}\}_{\mbox{\tiny\boldmath$\cal B$}^s}
+\{c,\{a,b\}_{\mbox{\tiny\boldmath$\cal B$}^s}\}_{\mbox{\tiny\boldmath$\cal B$}^s}
+\{b,\{c,a\}_{\mbox{\tiny\boldmath$\cal B$}^s}\}_{\mbox{\tiny\boldmath$\cal B$}^s}=0
\label{eq:jacobi}
\end{equation}
is satisfied as an identity  in canonical coordinates:
\begin{equation}
\{x_i,x_j\}_{\mbox{\tiny\boldmath$\cal B$}^s}={\cal B}_{ij}^s\;.
\label{cancoord}
\end{equation}
It is well-known that Poisson brackets can be used to
realize infinitesimal contact transformations~\cite{goldstein,mccauley,arnold}.
In particular
\begin{equation}
\hat{T}_q=\{\ldots,p\}_{\mbox{\tiny\boldmath$\cal B$}^s}\delta q
\end{equation}
is the operator realizing infinitesimal translations along the $q$ axis
and
\begin{equation}
\hat{T}_p=-\{\ldots,q\}_{\mbox{\tiny\boldmath$\cal B$}^s}\delta p
\end{equation}
is the corresponding operator along the $p$ axis
(carrying infinitesimal changes of the generalized momenta).
As a matter of fact, it can be easily verified that
$\hat{T}_qa(x)=(\partial a/\partial q)\delta q$ and
$\hat{T}_pa(x)=(\partial a/\partial p)\delta p$.
It is also easy to verify that, 
because of the canonical relations in~(\ref{cancoord}),
translations along the axis of different generalized coordinates, positions
and momenta, commute. This means that canonical phase space is \underline{flat}
even if generalized coordinates are used and even if the Lagrangian
manifold from which one builds phase space is a Riemann manifold~\cite{mccauley}.

In order to highlight the novel features which come into play
in the formalism of the non-Hamiltonian case, we think it
useful  to sketch briefly the equilibrium statistical mechanics of Hamiltonian
canonical systems in a form that can be easily generalized to the non-canonical
and non-Hamiltonian cases and that, by clearly distinguishing between
the Jacobian ${\cal J}(x)$ and the Jacobian determinant $\tilde{\cal J}(t,t_0)$,
is also suited to discuss time translation invariance.

It is well known that the Liouville operator can be
introduced using Poisson brackets,
$i\hat{L}=\{\ldots,{\cal H}\}_{\mbox{\tiny\boldmath$\cal B$}^s}$. 
The Liouville equation for the statistical
distribution function in phase space is written as~\cite{balescu}:
\begin{equation}
\frac{\partial\rho(x)}{\partial t}=-i\hat{L}\rho(x)\;.
\label{liouvcan}
\end{equation}
In Equation~(\ref{liouvcan}), the function
$\rho(x)={\cal J}^{can}f(x)$ has been introduced, where $f(x)$ is the true distribution
function in phase space and ${\cal J}^{can}=1$ is the Jacobian of transformations
between canonical coordinates.
In the canonical case, the identity $\rho(x)=f(x)$ trivially follows;
however, this notation will turn
out to be convenient later.
By means of the Liouville operator one can introduce the propagator
$\exp[it\hat{L}]=\exp[it\{\ldots,{\cal H}\}_P]$ whose action is defined by
\begin{equation}
a(x(t))=\exp[it\hat{L}]a(x)\;,
\end{equation}
where $x$ without the time argument denotes its value at time zero.
Statistical averages are calculated through
\begin{equation}
\langle a(x)\rangle=\int dx\rho(x)a(x)\;,
\label{eq:avea}
\end{equation}
and correlation functions as
\begin{equation}
\langle a b(t)\rangle=\int dx\rho(x)a(x)\exp[it\hat{L}]b(x)\;.
\label{eq:corra}
\end{equation}
We recall that in both Eqs.~(\ref{eq:avea}) and~(\ref{eq:corra}) 
the averaging over phase space coordinates
is done by considering the coordinates calculated at the initial time.
Indeed, the law of motion which carries $x(0)$ into $x(t)$
is another transformation of coordinates
\begin{equation}
dx(0)=\tilde{\cal J}(t,0)dx(t)\;,
\end{equation}
where the Jacobian determinant satisfies the condition
\begin{equation}
\tilde{\cal J}(t,0)\equiv\vert\frac{\partial x(0)}{\partial x(t)}\vert=1
\;.
\end{equation}
At equilibrium
\begin{equation}
\rho_{eq}(t)=e^{-iLt}\rho_{eq}(t)=\rho_{eq}=0\;,
\end{equation}
and
\begin{eqnarray}
\langle a(x)\rangle
&=&\langle a(x(t))\rangle\;.
\label{eq:timetranslaave}
\end{eqnarray}
The derivation of an analogous result for the correlation
functions is more subtle and will permit us
to make later important distinctions with the non-Hamiltonian case.
In the Hamiltonian canonical case, we get from Eq.~(\ref{eq:corra}):
\begin{eqnarray}
\langle ab(t)\rangle&=&
\int dx\rho_{eq}(x)a(x)b(x(t))
\nonumber\\
&=&\int dx(t)\rho_{eq}(x(t))
\left[e^{-itL}a(x(t))\right]
b(x(t))
\nonumber\\
&=&
\int dx(t)\rho_{eq}(x(t)) a(x(t)) e^{iLt}b(x(t))\;.
\end{eqnarray}
We consider 
the special case in which
\begin{equation}
b(x(0))=\left\{c(x(0)),d(x(0))\right\}_{\mbox{\tiny\boldmath$\cal B$}^s}\;,
\end{equation}
where $c(x)$ and $d(x)$ are two arbitrary phase space functions.
It is easy to prove, that if the Jacobi relation~(\ref{eq:jacobi})
\emph{holds}, then
\begin{equation}
e^{itL}\left\{c(x(0)),d(x(0))\right\}_{\mbox{\tiny\boldmath$\cal B$}^s}
=\left\{c(x(t)),d(x(t))\right\}_{\mbox{\tiny\boldmath$\cal B$}^s}
\;.
\end{equation}
Hence, it follows:
\begin{eqnarray}
\langle a&(x)&\left\{c(x(t)),d(x(t))\right\}_{\mbox{\tiny\boldmath$\cal B$}^s}\rangle
\nonumber\\
&&=\int dx(t)\rho_{eq}(x(t)) a(x(t))
e^{itL}
\left\{c(x(t)),d(x(t))\right\}_{\mbox{\tiny\boldmath$\cal B$}^s}
\nonumber\\
&&=\int dx(t)\rho_{eq}(x(t)) a(x(t))
\left\{c(x(2t)),d(x(2t))\right\}_{\mbox{\tiny\boldmath$\cal B$}^s}
\nonumber\\
=&&\langle a(x(t))\left\{c(x(2t)),d(x(2t))\right\}_{\mbox{\tiny\boldmath$\cal B$}^s}\rangle
\;.
\label{eq:corrinvtransla}
\end{eqnarray}
Equation~(\ref{eq:corrinvtransla}) shows that, at equilibrium
and in the Hamiltonian canonical case,
correlation functions are invariant under time translation.
We remark that the Jacobi relation~(\ref{eq:jacobi})
is necessary to derive this result so that it can be easily 
foreseen that in the non-Hamiltonian case, where the Jacobi relation
no longer holds, this property of time-correlation functions
is no longer verified.

\subsection{Non-canonical Hamiltonian Phase space}
\label{sec:noncanham}

Let us consider a transformation of phase space coordinates
$z=z(x)$ such that the Jacobian
\begin{equation}
{\cal J}=\|\frac{\partial x}{\partial z}\|\neq 1\;.
\label{noncanJ}
\end{equation}
Coordinates $z$ are called non-canonical.
The Hamiltonian transforms as a scalar ${\cal H}(x(z))={\cal H}^{\prime}(z)$
and the equations of motion become~\cite{mccauley,morrison}
\begin{equation}
\dot{z}_m=\sum_{k=1}^{2n}{\cal B}_{mk}(z)\frac{\partial{\cal H}^{\prime}(z)}
{\partial z_k}\;,
\label{noncanham}
\end{equation}
where
\begin{equation}
{\cal B}_{mk}=\sum_{i,j=1}^{2n}\frac{\partial z_m}{\partial x_i}{\cal B}_{ij}^s
\frac{\partial z_k}{\partial x_j}\;.
\label{tensorB}
\end{equation}
Equations of motion can also be obtained by means of the variational
principle~\cite{pericolanti} which arises when applying the non-canonical transformation
of coordinates to the symplectic expression of the action given in
Eq.~(\ref{symplaction}).
One obtains the following form for the action
in non-canonical coordinates~\cite{pericolanti}:
\begin{equation}
{\cal A}=\int dt\left[\frac{1}{2}\sum_{i,j,m=1}^{2n}\frac{\partial x_i}{\partial z_m}
\dot{z}_m{\cal B}_{ij}^sx_j(z)-{\cal H}^{\prime}(z)\right]\;,
\end{equation}
on which the variation is to be performed on the $z$ coordinates
in order to obtain Eqs.~(\ref{noncanham}).
Poisson brackets become non-canonical brackets defined by
\begin{equation}
\{a^{\prime},b^{\prime}\}
=\sum_{i,j=1}^{2n}\frac{\partial a^{\prime}(z)}{\partial z_i}{\cal B}_{ij}(z)
\frac{\partial b^{\prime}(z)}{\partial z_j}\;,
\label{noncanbrack}
\end{equation}
where $a^{\prime}(z)=a(x(z))$.
Non-canonical equations of motion can be expressed by means of the
bracket in Eq.~(\ref{noncanbrack}) as 
$\dot{z}_i=\{z_i,{\cal H}^{\prime}(z)\}_{\mbox{\tiny \boldmath$\cal B$}}$.
With a little bit of algebra, one can verify that non-canonical brackets
satisfy the Jacobi relation as an identity.
The Jacobi relation leads to the following identity
for $\mbox{\boldmath${\cal B}$}(z)$:
\begin{eqnarray}
S_{ijk}(z)&=&\sum_{l=1}^{2n}\left(
{\cal B}_{il}\frac{\partial{\cal B}_{jk}(z)}{\partial z_l}
+{\cal B}_{kl}\frac{\partial{\cal B}_{ij}(z)}{\partial z_l}
\right.\nonumber\\
&+&\left.{\cal B}_{jl}\frac{\partial{\cal B}_{ki}(z)}{\partial z_l}
\right) =0\;.
\label{sijk}
\end{eqnarray}
The non-canonical brackets of phase space coordinates are given by
\begin{equation}
\{z_i,z_j\}_{\mbox{\tiny\boldmath$\cal B$}}={\cal B}_{ij}(z)\;.
\label{noncancoord}
\end{equation}
Upon identifying the phase space point as
$z\equiv(\xi,\zeta)$,
one can define the operators
\begin{eqnarray}
\hat{T}_{\xi}&=&\{\ldots,\zeta\}_{\mbox{\tiny\boldmath$\cal B$}}\delta\xi\;,\\
\hat{T}_{\zeta}&=&-\{\ldots,\xi\}_{\mbox{\tiny\boldmath$\cal B$}}\delta\zeta\;,
\end{eqnarray}
which realize infinitesimal translations along the axes $\xi$ and $\zeta$.
The non-canonical bracket relations
of Eq.~(\ref{noncancoord}) imply that translations along the phase space axis
in general do not commute or, in other words, phase space is curved.
Notice we do not need to introduce a metric tensor. One just has
to introduce parallel transport and an affine connection,
which can be implicitly defined by means of the non-canonical brackets
and of the infinitesimal translation operators
$\hat{T}_{\xi}$ and $\hat{T}_{\zeta}$.

Non-canonical phase spaces are obtained by means of a non-canonical
transformation of coordinates applied to canonical Hamiltonian systems.
Suppose that there is a system with a non-canonical bracket satisfying the
Jacobi relation or its equivalent form given in Eq.~(\ref{sijk}).
Suppose also that $det\mbox{\boldmath${\cal B}$}\neq 0$. Then, by means of
Darboux's theorem the system can be put (at least locally) in a canonical form.
One would classify such a system as Hamiltonian:
following Refs.~\cite{mccauley,morrison}, we think that 
this property follows from the validity of the Jacobi relation.
In other words, if the algebra of brackets
is a Lie algebra, then the phase space is Hamiltonian~\cite{morrison}.

For non-canonical systems, statistical mechanical averages can be calculated as
\begin{equation}
\langle a^{\prime}(z)\rangle=\int dz \rho(z)a^{\prime}(z(t))\;,
\label{ncavg}
\end{equation}
where
\begin{equation}
\rho(z)={\cal J}(z) f^{\prime}(z)\;,
\end{equation}
the Jacobian ${\cal J}(z)$ being defined in Eq.~(\ref{noncanJ}).
The Liouville operator is defined by means of the non-canonical bracket
\begin{equation}
iL^{\prime}=\{\ldots,{\cal H}^{\prime}(z)\}_{\mbox{\tiny\boldmath$\cal B$}}\;,
\end{equation}
while time propagation is given by
\begin{eqnarray}
a^{\prime}(z(t))&=&\exp\left[itL^{\prime}\right]a^{\prime}(z)
\nonumber \\
&=&\exp\left[it\{\ldots,{\cal H}^{\prime}(z)\}_{\mbox{\tiny\boldmath$\cal B$}}\right]a^{\prime}(z)
\;.
\end{eqnarray}
In non-canonical coordinates a compressibility
\begin{equation}
\kappa(z)=\sum_{i,j=1}^{2n}\frac{\partial{\cal B}_{ij}(z)}{\partial z_i}
\frac{\partial{\cal H}^{\prime}(z)}{\partial z_j}
\end{equation}
\underline{might}, but not necessarily,
be present (see Appendix~\ref{app:1} for a simple example
of a non-canonical system with zero compressibility).
Integrating by parts Eq.~(\ref{ncavg}), one obtains
\begin{equation}
\langle a^{\prime}(z)\rangle=\int dz~
a^{\prime}(z)
\exp[-t(iL^{\prime}+\kappa(z))]\rho(z)
\;.\label{ncavgbyparts}
\end{equation}
Equation~(\ref{ncavgbyparts}) implies that $\rho(z)$ obeys the
non-canonical Liouville equation
\begin{eqnarray}
\frac{\partial\rho(z)}{\partial t}&=&-(iL^{\prime}+\kappa(z))\rho(z)
\nonumber\\
&=&\sum_{i=1}^{2n}\frac{\partial}{\partial z_i}(\dot{z}_i\rho(z))
\;.
\end{eqnarray}
It is easy to see that $d{\cal M}(z)={\cal J}(z)dz$
provides the correct invariant measure~\cite{pericolanti}:
\begin{eqnarray}
dz(t)\|\frac{\partial x(t)}{\partial z(t)}\|
&=&\|\frac{\partial z(t)}{\partial z(0)}\|dz(0)
\|\frac{\partial x(t)}{\partial x(0)}\|
\|\frac{\partial x(0)}{\partial z(t)}\|
\nonumber\\
&=&dz(0)\|\frac{\partial x(0)}{\partial z(0)}\|
\;,
\end{eqnarray}
where we have used the fact that the phase space flow
in the $x$ coordinates is canonical
so that $\|\partial x(t)/\partial x(0)\|=1$.
The use of the Jacobian provides the correct way of defining
the invariant measure because it can also be applied when
there is no compressibility~\cite{pericolanti}.
However, the calculation of averages, correlation functions
and linear response theory does not require the invariant measure explicitly.
The Knowledge of $\rho(z)$ is enough for statistical
mechanics even in the non-Hamiltonian case
and in the presence of constraints,
as shown in Refs.~\cite{bsiluranti,b1-2,bdispettose}.

It is interesting to analyze the time-translation invariance
of non-canonical statistical mechanics.
Under time evolution, the phase space volume element transforms as 
\begin{equation}
dz(0)=\tilde{\cal J}(t,0)dz(0)\;,
\end{equation}
where the Jacobian determinant
\begin{equation}
\tilde{\cal J}(t,0)=\vert\frac{\partial z(0)}{\partial z(t)}\vert
\label{eq:timejacobian}
\end{equation}
is different from unity \emph{if and only if}
the compressibility $\kappa$ is different from zero.
As the simple example developed in Appendix~\ref{app:1} shows,
one can also have non-canonical equations of motion with zero
compressibility. However, for the sake of comparison
with the papers cited in Ref.~\cite{tuck}, we shall explicitly consider
the case in which $\kappa\neq 0$ so that
$\tilde{\cal J}(t,0)=-\int_0^tdt'\kappa(t')$.
In this case and at equilibrium
\begin{eqnarray}
\langle a(z(0))\rangle
&=&\int dz(0)\rho_{eq}(z(0))a(z(0))
\nonumber\\
&=&\int dz(t)\tilde{\cal J}(t,0)
\rho_{eq}(z(0))
\left[e^{-itL}a(z(t))\right]\;.
\nonumber\\
\label{eq:aveazpart}
\end{eqnarray}
The law of evolution of the Jacobian in Eq.~(\ref{eq:timejacobian}) is:
\begin{equation}
\frac{d}{dt}\ln \left(\tilde{\cal J}(t,0)\right)=-\kappa(t)\;.
\label{eq:ddtlnJ}
\end{equation}
Now, let
\begin{equation}
\frac{d w(x(t))}{dt}=\kappa(x(t))\;.
\end{equation}
From Eq.~(\ref{eq:ddtlnJ}) it can be seen
that the function $w$, which is the primitive function
of $\kappa$, certainly exists.
Then, the Jacobian determinant can be written as
\begin{equation}
J(t,0)=e^{-w(x(t))}e^{w(x(0))}\;.
\label{eq:Jexplicit}
\end{equation}
The equilibrium distribution function follows as
\begin{equation}
\rho_{eq}(z(0))=Z^{-1}e^{-w(0)}\delta\left({\cal H}(z(0))-{\cal C}\right)\;
\label{eq:rhoeq}
\end{equation}
Then, substituting Eqs.~(\ref{eq:rhoeq}) and~(\ref{eq:Jexplicit})
into Eq.~(\ref{eq:aveazpart}), we obtain:
\begin{eqnarray}
&&\langle a(z(0))\rangle\nonumber\\
&&=\int dz(t)e^{-w(x(t))}
\frac{\delta\left({\cal H}(z(0))-{\cal C}\right)}{Z}
\left[e^{-itL}a(z(t))\right]
\nonumber\\
&=&\int dz(t)e^{-w(x(t))}
\frac{\delta\left({\cal H}(z(t))-{\cal C}\right)}{Z}
\left[e^{-itL}a(z(t))\right]\;,
\nonumber\\
\label{eq:aveazpart2}
\end{eqnarray}
where the second equality follows because the Hamiltonian is conserved
under time evolution. Then, upon integrating by parts, one obtains:
\begin{eqnarray}
\langle a(z(0))\rangle
&=&\int dz(t)\rho_{eq}(t)a(z(t))=\langle a(z(t))\rangle\;.
\label{eq:aveazpart3}
\end{eqnarray}
Equation~(\ref{eq:aveazpart3}) shows that phase space
averages are time-translation invariant in the non-canonical
statistical mechanics of systems at equilibrium.
Consider now the time correlation function
\begin{eqnarray}
\langle a\left\{b(t),c(t)\right\}_{\mbox{\tiny\boldmath$\cal B$}}\rangle
&=&
\int dz(0)\rho_{eq}(0)a(z(0))
\left\{b(t),c(t)\right\}_{\mbox{\tiny\boldmath$\cal B$}}
\nonumber\\
&=&
\int dz(t)e^{-w(t)}e^{w(0)}
e^{-w(0)}\delta\left({\cal H}(0)-{\cal C}\right)
\nonumber\\
&\times&
e^{-iLt}a(z(t))
\left\{b(t),c(t)\right\}_{\mbox{\tiny\boldmath$\cal B$}}
\nonumber\\
&=&
\int dz(t)\rho_{eq}(t)
a(z(t))e^{iLt}
\left\{b(t),c(t)\right\}_{\mbox{\tiny\boldmath$\cal B$}}
\nonumber\\
&=&
\int dz(t)\rho_{eq}(t)
a(z(t))
\left\{b(2t),c(2t)\right\}_{\mbox{\tiny\boldmath$\cal B$}}
\nonumber\\
&=&
\langle a(t)\left\{b(2t),c(2t)\right\}_{\mbox{\tiny\boldmath$\cal B$}}\rangle
\;,\label{eq:nccorrtimetransla}
\end{eqnarray}
where the last equality arises again from the validity of the Jacobian
relation~(\ref{eq:jacobi}).
Equation~(\ref{eq:nccorrtimetransla}) shows that,
when the bracket satisfies the Jacobi relation,
the equilibrium statistical mechanics is invariant
under time translation.
We would like to recall that an
algebra expressed by brackets 
satisfying the Jacobi relation is a Lie algebra.
Therefore,
it is reasonable to consider a phase space theory 
observing such a property as Hamiltonian.

\subsection{Non-Hamiltonian Phase Space}
\label{sec:nonham}

In Refs.~\cite{b1-2,bdispettose} it was shown how non-Hamiltonian
equations of motion, brackets and statistical mechanics
can be defined.
One must simply keep the generalized symplectic structure
of the non-canonical equations in Eq.~(\ref{noncanham})
and of the non-canonical bracket in Eq.~(\ref{noncanbrack})
and use, in place of $\mbox{\boldmath${\cal B}$}(z)$,
an antisymmetric matrix $\tilde{\mbox{\boldmath${\cal B}$}}(z)$.
The matrix $\tilde{\mbox{\boldmath${\cal B}$}}(z)$ can be chosen
arbitrarily, with the only constraint of being antisymmetric
so that the Hamiltonian is conserved. As a result the Jacobi relation
may not be satisfied~\cite{b1-2}.
When this happens, one of the conditions
of validity of the Darboux's theorem fails
so that non-Hamiltonian phase space flows cannot be
put into a canonical form.
The failure of the Jacobi relation implies
\begin{equation}
e^{iLt}\left\{a(0),b(0)\right\}_{\tilde{\mbox{\tiny\boldmath$\cal B$}}}
\neq 
\left\{a(t),b(t)\right\}_{\tilde{\mbox{\tiny\boldmath$\cal B$}}}
\;.\label{eq:notimetransla}
\end{equation}
Hence, even if the non-Hamiltonian theory has a vanishing compressibility,
the
equilibrium statistical mechanics is not time translation invariant.
To show this, we note that:
\begin{eqnarray}
\langle
\left\{a(0),b(0)\right\}_{\tilde{\mbox{\tiny\boldmath$\cal B$}}}
\rangle
&=&
\int d\tilde{z}(0)\tilde{\rho}_{eq}(0)
\left\{a(0),b(0)\right\}_{\tilde{\mbox{\tiny\boldmath$\cal B$}}}
\nonumber\\
&\neq&
\int d\tilde{z}(t)\tilde{\rho}_{eq}(t)
\left\{a(t),b(t)\right\}_{\tilde{\mbox{\tiny\boldmath$\cal B$}}}
\end{eqnarray}
because of Eq.~(\ref{eq:notimetransla}).

We consider the lack of time translation invariance to be the crucial
feature of non-Hamiltonian statistical mechanics.

\subsubsection{Non-Hamiltonian equations of motion with 
no phase space compressibility}

It is easy to realize that there are non-Hamiltonian
phase space flows (\emph{i.e.}, flows defined by brackets which do not satisfy
the Jacobi relation) with zero compressibility.
We refer the reader to Appendix~\ref{app:2} for a trivial example.
In order to show how this is possible,
we consider a particular sub-ensemble of non-Hamiltonian
phase space flows, \emph{viz.}, those flows that can be derived by means
of a non-integrable scaling of time.
Interestingly, it was Nos\'e who originally considered
this type of flows when he introduced his
celebrated thermostat~\cite{nose,ferrario}.
Nos\'e started with a canonical Hamiltonian system and then performed
a non-canonical transformation, followed by a non-integrable
scaling of time. Accordingly, we 
consider the non-integrable scaling of time
\begin{equation}
dt=\Phi(z)d\tau\;,
\end{equation}
where $\tau$ is an auxiliary time variable.
Such a scaling of $dt$ is clearly non-integrable because,
due to the dependence of $dt$ on phase space coordinates,
the integral $\int dt$ depends on the path in phase space.
If we apply this scaling to Eq.~(\ref{noncanham}),
we obtain the following non-Hamiltonian
equations:
\begin{eqnarray}
\frac{d z_i}{d\tau}&=&\sum_{j=1}^{2n}\Phi(z){\cal B}_{ij}(z)
\frac{\partial{\cal H}^{\prime}(z)}{\partial z_j}
\nonumber\\
&=&
\sum_{j=1}^{2n}\tilde{\cal B}_{ij}(z)
\frac{\partial{\cal H}^{\prime}(z)}{\partial z_j}
\;.
\label{nonhameq}
\end{eqnarray}
Using the antisymmetric matrix $\mbox{\boldmath$\tilde{\cal B}$}$ as
defined in Eq.~(\ref{nonhameq}), one can introduce
a non-Hamiltonian bracket
\begin{equation}
\left\{a^{\prime},b^{\prime}\right\}_{\tilde{\mbox{\tiny\boldmath$\cal B$}}}
=\sum_{i,j=1}^{2n}
\frac{\partial a^{\prime}}{\partial z_i}\tilde{\cal B}_{ij}(z)
\frac{\partial b^{\prime}}{\partial z_j}
\;.
\label{nonhambrack}
\end{equation}
This bracket does not satisfy the Jacobi relation
so that the
equations of motion~(\ref{nonhameq}) are non-Hamiltonian.
In this case, there is a non-vanishing tensor
associated with the Jacobi relation:
\begin{eqnarray}
\tilde{S}_{ijk}
&=&\sum_{l=1}^{2n}\left(
\tilde{\cal B}_{il}(z)\frac{\partial\tilde{\cal B}_{jk}(z)}{\partial z_l}
+\tilde{\cal B}_{kl}(z)\frac{\partial\tilde{\cal B}_{ij}(z)}{\partial z_l}
\right.\nonumber\\
&+&\left.\tilde{\cal B}_{jl}(z)\frac{\partial\tilde{\cal B}_{ki}(z)}{\partial z_l}
\right)
\neq 0\;.
\end{eqnarray}
The compressibility of the non-Hamiltonian equations~(\ref{nonhameq})
is given by
\begin{eqnarray}
\tilde{\kappa}(z)&=&\sum_{i=1}^{2n}
\frac{\partial (d z_i/d\tau)}{\partial z_i}
=\sum_{i,j=1}^{2n}\frac{\partial\tilde{\cal B}_{ij}}{\partial z_i}
\frac{\partial{\cal H}^{\prime}(z)}{\partial z_j}
\nonumber\\
&=&\sum_{i=1}^{2n}\frac{\partial\Phi(z)}{\partial z_i}\frac{d z_i}{dt}
+\Phi(z)\kappa(z)\;,
\end{eqnarray}
where $dz_i/dt$ and $\kappa$ are given by the non-canonical equations of motion
prior to the non-integrable time-scaling.
It is evident that whenever one chooses $\Phi(z)$ in such a way that
\begin{equation}
\sum_{i=1}^{2n}\frac{\partial\ln\Phi(z)}{\partial z_i}
\frac{dz_i}{dt}=-\kappa(z)\;,
\label{kappacondition}
\end{equation}
non-Hamiltonian flows  have vanishing compressibility
(see the trivial example given in Appendix~\ref{app:2}).
Correspondingly, when the scaling is chosen according to Eq.~(\ref{kappacondition}),
the distribution function will obey a non-Hamiltonian equation 
without a compressibility:
\begin{eqnarray}
\frac{\partial \tilde{\rho}(z)}{\partial \tau}
&=&-\left\{\tilde{\rho},{\cal H}^{\prime}(z)\right\}_{\tilde{\mbox{\tiny\boldmath$\cal B$}}}
=-i\tilde{L}\tilde{\rho}(z)\;.
\end{eqnarray}

\section{Relative entropy}\label{sec:relativeentropy}

\subsection{The Hamiltonian case}\label{sec:hamentropy}

As is well known, the definition of the entropy functional
for systems with continuous probability distribution,
needs a special care. We review, for the reader's convenience,
some relevant definitions which hold for the Hamiltonian case
and which we plan to generalize to non-Hamiltonian systems
in the following sections.

In order to be rigorous, one first assumes that phase space
can be divided into small cells of volume $\Delta^{(i)}$ so that
the coordinates in the $i$-th cell are denoted as $x^{(i)}$.
This way, phase space is effectively discretized and so
is the distribution function: $\rho(x^{(i)})\equiv\rho^{(i)}$.
The absolute information entropy can be defined as~\cite{rau,jaynes}:
\begin{equation}
S[\rho]=-k_B\sum_i\rho^{(i)}\ln\rho^{(i)}\;,
\end{equation}
where $k_B$ is Boltzmann's constant.
The continuous limit
\begin{equation}
S[\rho]=\lim_{\Delta^{(i)}\to 0}
\left(-k_B\sum_i\rho^{(i)}\ln\rho^{(i)}\right)\;,
\end{equation}
diverges.
Upon subtracting the divergent contribution, $-k\ln\Delta^{(i)}$,
one obtains the finite expression
\begin{equation}
S[\rho]
=-k_B\int dx\rho(x)\ln\rho(x)\;.\label{snaiveprime}
\end{equation}
However, as discussed in Refs.~\cite{stratonovich,ramshaw,rau},
the definition given in Eq.~(\ref{snaiveprime}) is not acceptable
because it is not coordinate independent.

When studying continuous probability distributions,
one should use the \underline{relative} entropy,
which is a measure of the information~\cite{rau} relative to a state of ignorance
(represented by a given distribution function).
If one denotes this latter distribution by $\mu(x)$,
the relative entropy reads as
\begin{equation}
S_{rel}[\rho]=-k_B\int dx\rho(x)\ln\left(\frac{\rho(x)}{\mu(x)}\right)
\;.\label{srel}
\end{equation}
For canonical Hamiltonian systems, a convenient distribution
function with respect to which one can define the relative entropy,
is given by the distribution representing the state of absolute ignorance,
\emph{i.e.}, the uniform distribution. Then, one can set $\mu(x)=1$, so that,
in practice, $S_{rel}[\rho]$ in Eq.~(\ref{srel}) coincides
with the absolute entropy $S[\rho]$ in Eq.~(\ref{snaiveprime}).
However, it must be realized that a uniform distribution
in canonical coordinates does not necessarily transform into
another uniform distribution if more general coordinates (for example non-canonical)
are used.
Instead, the integral in Eq.~(\ref{srel}) is well defined and its value does not
depend on a specific choice of coordinates.
For example, considering the transformation $x\to y$, one has:
\begin{equation}
\int dx\rho(x)\ln\left(\frac{\rho(x)}{\mu(x)}\right)
=
\int dy\rho^{\prime}(y)\ln\left(\frac{\rho^{\prime}(y)}{\mu^{\prime}(y)}\right)
\;,
\end{equation}
where $dx\rho(x)=dy\rho^{\prime}(y)$ and $dx\mu(x)=dy\mu^{\prime}(y)$.

The relative entropy $S_{rel}[\rho]$ in Eq.~(\ref{srel}) is a measure
of missing information and can be used as the starting point
of a maximum entropy principle in order to obtain the form of
the least biased or maximum non-committal distribution function $\rho(x)$.
To this end, upon considering the two statistical constraints
$\langle{\cal H}\rangle=E$  and $\int dx \rho(x)=1$,
one is led to consider the quantity
\begin{equation}
{\cal I}=
S_{rel}[\rho]+\lambda\left(E- \langle{\cal H}\rangle\right)+\gamma\left(1-\int dx\rho(x)\right)
\;,
\end{equation}
where two Lagrangian multipliers ($\lambda$ and $\gamma$) have been introduced.
Upon maximizing ${\cal I}$ with respect to $\rho(x)$,
the following expression for $\rho(x)$ is easily recovered:
\begin{equation}
\rho(x)=Z^{-1}\mu(x)\exp\left[-\frac{\lambda}{k_B}{\cal H}(x)\right]\;,
\label{rhomaxen}
\end{equation}
where $Z=\int dx \mu(x)\exp[-(\lambda/k_B){\cal H}(x)]$.
Equation~(\ref{rhomaxen}) generalizes
the standard maximum entropy principle~\cite{jaynes}.
to the case of the relative entropy of continuous probability distributions.
In the canonical Hamiltonian case, which has been treated in this section, $\mu(x)$
is trivially the uniform distribution. However, for non-canonical and non-Hamiltonian
phase space $\mu(x)$ plays a fundamental role.

\subsection{Non-canonical and non-Hamiltonian
relative entropy}\label{sec:ncrelentropy}

In the non-canonical and non-Hamiltonian case
a Jacobian enters the definition of the phase space
volume element.
Accordingly, one should write
a coordinate-independent entropy functional  as
\begin{eqnarray}
S_{\cal J}&=&-k_B\int dz{\cal J}(z)f^{\prime}(z)
\ln f^{\prime}(z)\nonumber\\
&=&-k_B\int dz\rho(z)\ln\left(\frac{\rho(z)}{{\cal J}(z)}\right)
\;.\label{eq:recover}
\end{eqnarray}
This expression is recovered starting from the relative entropy.
An intrinsic (coordinate-free) form of Eq.~(\ref{eq:recover})
has been given in Ref.~\cite{ezra2}.
In order to see this, one just needs to transform Eq.~(\ref{srel})
into non-canonical coordinates: 
\begin{eqnarray}
\mu(x)dx&=&1\cdot dx={\cal J}(z)dz\\
f(x)dx&=& f^{\prime}(z){\cal J}(z)dz=\rho(z)dz\;.
\end{eqnarray}
so that Eq.~(\ref{eq:recover}) naturally follows.
If the Jacobian ${\cal J}(z)$ is not known, one can use any other
distribution function with respect to which the entropy is calculated,
\emph{i.e.}, $m(x)dx=m^{\prime}(z){\cal J}(z)dz$. Let us define
$\mu(z)=m^{\prime}(z){\cal J}(z)$, in analogy with
$\rho(z)=f^{\prime}(z){\cal J}(z)$.
One can think of $\mu(z)$ as the solution of a Liouville equation
with different interactions. For example, if
$iL^{\prime}=iL^{\prime}_0+iL^{\prime}_I$, one may define
$\mu(z)$ as the solution of the equation
\begin{equation}
\frac{\partial\mu(z)}{\partial t}
=-(iL^{\prime}_0+\kappa_0)\mu(z)\;,
\end{equation}
with $\kappa=\kappa_0+\kappa_I$.
The entropy determined by the additional interactions, represented by
$iL^{\prime}_I$, with respect to the state where the term $iL^{\prime}_I$
is absent, is given by
\begin{equation}
S_{rel}[\rho|\mu]=-k_B\int dz\rho(z)\ln\left(\frac{\rho(z)}{\mu(z)}\right)
\;.\label{srelnc}
\end{equation}
Equation~(\ref{srelnc}), with the correct interpretation
of the distribution function $\mu(z)$, provides a coordinate-invariant
definition of the relative entropy which does not require the explicit
knowledge of the \emph{metric} as well as of the Jacobian.
The maximum-entropy principle, as written in the previous section for
the canonical case, also applies without major changes
to the non-canonical and non-Hamiltonian dynamics.
The functional to be maximized is
\begin{eqnarray}
{\cal I}=S_{rel}[\rho|\mu]
+\lambda\left(E-\langle{\cal H}^{\prime}(z)\rangle\right)
+\gamma\left(1-\int dz\rho(z)\right)\;,
\nonumber
\end{eqnarray}
which provides, by setting $\delta {\cal I}/\delta\rho(z)=0$,
the generalized canonical distribution in non-canonical
coordinates
\begin{equation}
\rho(z)=Z_{\mu}^{-1}\mu(z)\exp
\left[-\frac{\lambda}{k_B}{\cal H}^{\prime}(z)\right]
\;.
\end{equation}
The quantity
$Z_{\mu}=\int dz \mu(z)\exp[-(\lambda/k_B){\cal H}^{\prime}(z)]$
is a \emph{weighted} partition function.

\subsection{Relative entropy production}
\label{sec:enprod}

In order to simplify the notation, we rewrite the Liouville equation as
\begin{equation}
\frac{\partial\rho(x,t)}{\partial t}=-\nabla\cdot(\rho\dot{x})\;,
\end{equation}
where we have introduced an obvious vectorial notation (to avoid indices)
and $\nabla=\partial/\partial x$ is the operator of differentiation
with respect to phase space coordinates.
From the relative entropy functional
\begin{equation}
S=-k_B\int dx\rho\ln\left(\gamma^{-1}\rho\right)\;,
\end{equation}
the entropy production follows as
\begin{eqnarray}
\dot{S}
&=&-k_B\int dx\left[-\nabla\cdot(\rho\dot{x})\ln(\gamma^{-1}\rho)
\right.\nonumber\\
&+&\left.\gamma\left( \gamma^{-1}\partial_t\rho
-\rho\gamma^{-2}\partial_t\gamma \right)\right]
\;.
\end{eqnarray}
The term $\int dx\partial_t\rho$ vanishes because of the normalization condition
on $\rho$. Hence
\begin{eqnarray}
\dot{S}
&=&-k_B\int dx\left[\rho\dot{x}\cdot\nabla\ln(\gamma^{-1}\rho)
- \rho\gamma^{-1}\partial_t\gamma \right]\;.
\end{eqnarray}
The entropy production can then be put in the form
\begin{eqnarray}
\dot{S}
&=&
k_B\langle \gamma^{-1}\partial_t\gamma\rangle
-k_B\int dx\rho\dot{x}\cdot
\left(\frac{\nabla\rho}{\rho}-\frac{\nabla\gamma}{\gamma}\right)
\;.\label{eq:relentrprod1}
\end{eqnarray}

We can show that Eq.~(\ref{eq:relentrprod1}) coincides
with the formula of the covariant entropy production formerly
given by Ramshaw~\cite{ramshaw} when $\gamma=\rho f^{-1}$.
To this end, upon noting that
\begin{eqnarray}
\rho^{-1}\nabla\rho&=& f^{-1}\nabla f+\gamma^{-1}\nabla\gamma
\end{eqnarray}
we obtain:
\begin{eqnarray}
\int dx\rho\dot{x}\cdot\big[\frac{\nabla\rho}{\rho}
-\frac{\nabla\gamma}{\gamma}\big]
&=&-\int dx \rho\gamma^{-1}\nabla\cdot(\gamma\dot{x}) \;.
\end{eqnarray}

In the general case, we obtain from Eq.~(\ref{eq:relentrprod1})
\begin{eqnarray}
\dot{S} &=&
k_B\langle\frac{1}{\gamma}
\left(\frac{\partial\gamma}{\partial t}
+\dot{x}\cdot\nabla\gamma\right)\rangle
+k_B\langle\kappa\rangle \;,
\end{eqnarray}
and upon noting that
\begin{eqnarray}
\langle\kappa\rangle
&=&\int dx\rho\frac{1}{\gamma}\left(\gamma\nabla\cdot \dot{x}\right)\;.
\end{eqnarray}
the relative entropy production can be written as
\begin{equation}
\dot{S}=k_B\langle\omega(x,t)\rangle\;,\label{eq:relentrprodf}
\end{equation}
where 
\begin{equation}
\omega(x,t)=\frac{1}{\gamma}\left[\frac{\partial\gamma}{\partial t}
+\nabla\cdot(\gamma\dot{x})\right]\;.
\end{equation}
Equation~(\ref{eq:relentrprodf}) is identical to
the covariant entropy production given by Ramshaw~\cite{ramshaw}.
Ezra~\cite{ezra2} has also provided a coordinate-free expression of
Eq.~(\ref{eq:relentrprodf}). Such comparisons are presented to validate
our information theoretical approach to non-Hamiltonian systems at equilibrium.
When $\omega=0$, \emph{i.e.}, if $\gamma$ is a solution of the Liouville
equation, then $\dot{S}=0$. It is also clear that in an equilibrium ensemble
($\partial_t\rho=0$, $\partial_t f=0$, $\partial_t\gamma=0$)
the production of (relative) entropy is trivially null.

\section{Conclusions} \label{sec:conclusions}

The geometry of phase space is peculiar. 
Antisymmetric brackets, which define a Lie algebra (in the canonical
and non-canonical cases) or a non-Lie algebra (in the non-Hamiltonian case),
can be used to connect, by means of infinitesimal ``contact'' transformations,
nearby points over the manifold.
A distinctive feature of non-Hamiltonian
equilibrium statistical mechanics is the lack of time-translation invariance.
This is mathematically represented by the failure of the Jacobi relation.
As a matter of fact, whenever a bracket satisfies the Jacobi relation,
the equilibrium statistical mechanics is time translation invariant
and the bracket realizes a Lie algebra.
We surmise that such theories  should be classified as Hamiltonian.

We argue that a non-vanishing phase space compressibility is not a signature
of non-canonical or non-Hamiltonian dynamics. There are cases
(and it is worth noting that Andersen's constant pressure 
dynamics~\cite{andersen} is one of these) 
where the compressibility is zero
but the Jacobian is not unity and
the dynamics is non-canonical (or non-Hamiltonian).
In particular,  we have derived
a condition for having a vanishing compressibility
in a restricted class
of non-Hamiltonian phase space flows (those flows
which are obtained by means of non-canonical transformations 
of coordinates followed
by a non-integrable scaling of time~\cite{nose,ferrario}).
In such cases, it is obvious that the measure of phase space cannot be 
derived from the phase space compressibility.
Instead, one should use the Jacobian of the transformation
between different types of phase space coordinates.
However, it is not necessary to know explicitly such a Jacobian 
for setting up a statistical mechanical theory. In fact,
the distribution function and the Liouville operators
are  sufficient to this end.

We remark how, for continuous probability distributions,
one should use the relative entropy functional.
The definition of the relative entropy is coordinate independent
and, measuring the state of ignorance relative to another given distribution,
does not require the explicit knowledge of the Jacobian.
A maximum-entropy principle, which applies to the relative
entropy, has been formulated in the canonical, 
non-canonical and non-Hamiltonian case.
Such a maximum-entropy principle may turn out to be relevant 
for applications in the non-Hamiltonian dynamics 
of non-equilibrium thermodynamical ensembles.

\vspace{0.5cm}

\noindent
{\bf Acknowledgment}

\noindent
We thank Professor Gregory S. Ezra for his useful comments
and for a critical reading of the manuscript.


\appendix
\section{A non-canonical system with zero compressibility}
\label{app:1}

Consider the following simple Hamiltonian
\begin{equation}
{\cal H}=\frac{p_1^2}{2}+\frac{p_2^2}{2}
+\frac{1}{2}(q_1-q_2)^2\;,
\label{harmham}
\end{equation}
whose canonical equations of motion can be easily written down.
Consider, instead, the following non-canonical transformation
of coordinates $x=(q_1,q_2,p_1,p_2)
\to z=(\xi_1,\xi_2,\pi_1,\pi_2)$ defined by
\begin{eqnarray}
q_1&=&\xi_1\xi_2^{-1}\\
q_2&=&\xi_2\\
p_1&=&\xi_2\pi_1\\
p_2&=&\pi_2\;.
\end{eqnarray}
By using this transformation of coordinates onto the
canonical equation of motion, one obtains the following non-canonical
equations of motion
\begin{eqnarray}
\dot{\xi}_1&=&\xi_1\pi_2\xi_2^{-1}+\xi_2^2\pi_1
\label{eqapp1_1}\\
\dot{\xi}_2&=&\pi_2\\
\dot{\pi}_1&=&-\pi_2\xi_2^{-1}\pi_1+\xi_2^{-1}(\xi_2-\xi_1\xi_2^{-1})\\
\dot{\pi}_2&=&-(\xi_2-\xi_1\xi_2^{-1})\;.
\label{eqapp1_2}
\end{eqnarray}
The Hamiltonian in non-canonical coordinates is
\begin{equation}
{\cal H}^{\prime}(z)=\xi_2^2\frac{\pi_1^2}{2}
+\frac{\pi_2^2}{2}+\frac{1}{2}(\xi_1\xi_2^{-1}-\xi_2)^2\;.
\end{equation}
One can calculate
\begin{eqnarray}
\frac{\partial{\cal H}^{\prime}(z)}{\partial\xi_1}
&=&\xi_2^{-1}(\xi_1\xi_2^{-1}-\xi_2)
\\
\frac{\partial{\cal H}^{\prime}(z)}{\partial\xi_2}
&=&\xi_2\pi_1^2-(\xi_1\xi_2^{-1}-\xi_2)
(\xi_1\xi_2^{-2}+1)
\\
\frac{\partial{\cal H}^{\prime}(z)}{\partial\pi_1}
&=&a\xi_2^2\pi_1
\\
\frac{\partial{\cal H}^{\prime}(z)}{\partial\pi_2}
&=&\pi_2\;,
\end{eqnarray}
and write the equations in matrix form
\begin{eqnarray}
\left[\begin{array}{c}\dot{\xi}_1\\\dot{\xi}_2\\
\dot{\pi}_1\\\dot{\pi}_2\end{array}\right]
&=&
\left[\begin{array}{cccc}
0 & 0 & 1 & \xi_1\xi_2^{-1}\\
0 & 0 & 0 & 1 \\
-1 & 0 & 0 & -\pi_1\xi_2^{-1}\\
-\xi_1\xi_2^{-1} & -1 & \pi_1\xi_2^{-1} & 0
\end{array}\right]\nonumber\\
& & \nonumber\\
&\times&
\left[\begin{array}{c}
\partial{\cal H}^{\prime}(z)/\partial\xi_1
\\
\partial{\cal H}^{\prime}(z)/\partial\xi_2
\\
\partial{\cal H}^{\prime}(z)/\partial\pi_1
\\
\partial{\cal H}^{\prime}(z)/\partial\pi_2
\end{array}\right]\;.\label{eqappmat}
\end{eqnarray}
Equations~(\ref{eqapp1_1}-\ref{eqapp1_2})
are obviously non-canonical, as it is clearly seen
by their matrix form given in Eq.~(\ref{eqappmat}),
and they have zero compressibility.
Hence,
\emph{metric} theories cannot be applied.
The antisymmetric matrix appearing in Eq.~(\ref{eqappmat})
must be used to define the non-canonical bracket, which obviously
satisfies the Jacobi relation and the Liouville equation.

\section{A non-Hamiltonian system with zero compressibility}
\label{app:2}

Let us consider the Hamiltonian of Eq.~(\ref{harmham}).
We first obtain non-canonical equations of motion by considering
the transformation of coordinates
\begin{eqnarray}
q_1&=&\xi_2\xi_1\\
q_2&=&\xi_2\\
p_1&=&\pi_1\\
p_2&=&\pi_2\;.
\end{eqnarray}
The Hamiltonian becomes
\begin{equation}
{\cal H}^{\prime}
=\frac{\pi_1^2}{2}+\frac{\pi_2^2}{2}
+\frac{\xi_2^2}{2}(\xi_1-1)^2\;.
\end{equation}
The non-canonical equations of motion are
\begin{eqnarray}
\dot{\xi}_1&=&\xi_2^{-1}\pi_1-\xi_1\xi_2^{-1}\pi_2\\
\dot{\xi}_2&=&\pi_2\\
\dot{\pi}_1&=&-\xi_2(\xi_1-1)\\
\dot{\xi}_2&=&-\xi_2(1-\xi_1)\;.
\end{eqnarray}
Such non-canonical equations have a compressibility
\begin{equation}
\kappa=-\xi_2^{-1}\pi_2\;.
\end{equation}
We define the antisymmetric matrix
\begin{eqnarray}
\mbox{\boldmath${\cal B}$}
&=&\left[\begin{array}{cccc}
0 & 0 & \xi_2^{-1} & -\xi_2^{-1}\xi_1\\
0 & 0 & 0 & 1 \\
-\xi_2^{-1} & 0 & 0 & 0\\
\xi_2^{-1}\xi_1 & -1 & 0 & 0\end{array}\right]
\end{eqnarray}
which should be used in order to define the non-canonical bracket
which satisfies the Jacobi relation.
Now, if one wants to apply a non-integrable scaling of time
in order to obtain a non-Hamiltonian flow with zero compressibility
$\tilde{\kappa}$, Eq.~(\ref{kappacondition}) can be used.
Assuming the scaling function $\Phi=\Phi(\xi_2)$,
one obtains
\begin{equation}
\frac{\partial\Phi}{\partial\xi_2}\pi_2=\xi_2^{-1}\pi_2\;,
\end{equation}
from which one readily finds $\Phi=\xi_2$.
Hence, the antisymmetric matrix $\tilde{\mbox{\boldmath$\cal B$}}=\Phi
\mbox{\boldmath$\cal B$}$ is
\begin{eqnarray}
\tilde{\mbox{\boldmath${\cal B}$}}
&=&\left[\begin{array}{cccc}
0 & 0 & 1 & -\xi_1\\
0 & 0 & 0 & \xi_2 \\
-1 & 0 & 0 & 0\\
\xi_1 & -\xi_2 & 0 & 0\end{array}\right]
\;.\label{btildeexample}
\end{eqnarray}
Non-Hamiltonian equations of motion are now defined according to
Eq.~(\ref{nonhameq}). Finally, 
it is not difficult to verify that the non-Hamiltonian bracket
of Eq.~(\ref{nonhambrack}),
with $\tilde{\mbox{\boldmath$\cal B$}}$ defined
in Eq.~(\ref{btildeexample}), does not satisfy the Jacobi relation
and $\tilde{S}_{ijk}\neq 0$. For example, it is easy to verify that
$\tilde{S}_{314}=1$.


\end{document}